\documentclass[a4paper]{elsart}
\usepackage{graphicx}
\usepackage{tabularx}
\usepackage{rotating}
\usepackage{subfigure}
\begin{document}
\begin{frontmatter}
\begin{flushright}
DESY 08-047
\end{flushright}
\title{Study of Micro Pixel Photon Counter \\for the application to Positron Emission Tomography}
\author[AHHUNI,DESY]{N. D'Ascenzo\thanksref{email}},
\author[DESY]{E. Garutti},
\author[DESY]{M. Goettlich},
\author[HEIUNI]{H.C. Schultz-Coulon},
\author[HEIUNI]{A. Tadday}

\address[AHHUNI]{University of Hamburg, Edmund-Siemers-Allee 1, D-20146, Hamburg, Germany}
\address[DESY]{DESY, Notkestr. 85, D-22607 Hamburg, Germany}
\address[HEIUNI]{University of Heidelberg, Im Neuenheimer Feld 227,
D-69120 Heidelberg, Germany}
\thanks[email]{Corresponding author: nicola.dascenzo@desy.de}
\begin{abstract}
The main challenges posed by the design of future Positron Emission Tomography machines are the improvement of the spatial and timing resolution and the combined operation with magnetic resonance. The Micro Pixel Photon Counter by Hamamatsu is a good candidate for this application. Its small size (down to $1\times 1$~mm$^{2}$) and the high photo-detection efficiency in the blue spectral region allow the direct readout of a highly segmented scintillator matrix improving the spatial resolution of the machine. Furthermore, this photo-detector is insensitive to static magnetic fields up to 5~T, which makes it a possible candidate for applications in a magnetic resonance environment, though tests in a fast changing gradient field need still to be performed. 

The aim of this study is the characterization of a system composed by a scintillator crystal readout via a MPPC. Crystals of  $1\times 1 \times 15$~mm$^{3}$ and $3\times 3 \times 15$~mm$^{3}$ are directly coupled to a MPPC of the same size active area and the energy resolution at 511 keV is measured. The coincidence time resolution of two so assembled detector units is measured.  A first comparison of the performances of LSO and LFS crystals is given.
\end{abstract}
\end{frontmatter}

\section*{Introduction}
Positron Emission Tomography (PET) is a non-invasive medical imaging technique \cite{PET1,PET2}. A $\beta^{+}$ emitter is used to mark a tracer (generally a glucose molecule) which is injected into a living organism. 
The two 511-keV photons, produced by the e$^+$e$^-$ annihilation inside the organism, are detected in coincidence and their line of response is identified.
The tomographic reconstruction of several lines of response via well established mathematical techniques allows to retrieve the 3D image of the organs in which the tracer is absorbed.
The typical detector block for commercial PET consists of a pixelated BGO crystal, readout by up to 10 photomultiplier tubes. The signals from the photo-detectors are weighted by a resistive chain and the interaction point in the scintillator is reconstructed \cite{PET1}.

The energy resolution at the 511 keV photo-peak and the time response are two relevant parameters for the optimization of a PET system. The energy resolution makes it possible to identify and neglect photons which have undergone Compton scattering. Compton-scattered photons change direction and contribute to the reconstruction of fake lines of response. These events are regarded as background to the PET measurements. 
A standard BGO-based PET detector  provides $10-13\%$ energy resolution at 511 keV. 
The time resolution influences the width of the coincidence window
for the two photons and therefore the background caused by random coincidences. It is determined by the decay time of the scintillator and the response of the readout photo-detector. If BGO is used, the time resolution is mainly featured by the decay time of the scintillator, $\sim300$~ns. A shorter coincidence window ($\textsl{O}(10)$~ns) would be optimal, as it reduces the random coincident background. \\
The time information can also be used to directly improve the position resolution as done in the Time-of-Flight PET (TOF-PET). 
A time resolution of 500~ps FWHM corresponds to a localization of the source in an interval of 7.5~cm, resulting in the enhancement of the signal to noise ratio of a factor 2 \cite{Bill1,Bill2}. 

The spatial resolution is mainly determined by the pixel size  of the PET camera, as well as by the coupling between scintillator and photo-detector and by the reconstruction techniques used. Concerning the pixel size it has been shown that an array of crystals read out by an array of photo-detectors with the same granularity performs better than the usual light sharing blocks \cite{Lecomte}.   

Recent developments in PET technologies are mainly focused on the improvement of energy and time resolution of both the crystals as well as of the photo-detectors used. Many new fast crystals have been produced in the last ten years. LSO (Lutetium Orthosilicate, or Lu$_{2}$SiO$_{5}$) is one of them. It has  a decay time of 40~ns and emits a photon spectrum peaked at 420~nm wavelength. Even faster crystals are available, featured by a peak light emission in the blue and ultra-violet spectrum. The characteristics of these fast crystals dictate the requirements for the photo-detectors for PET. They must have a good photo-detection efficiency in the blue spectral range and an excellent time response. 

The Micro Pixel Photon Counter (MPPC) by Hamamatsu \cite{Hama} is an excellent candidate for this application.
It is a silicon photo-detector, with a design similar to the Silicon
Photomultiplier \cite{SIPM}, produced in variable size from
1x1~mm$^{2}$ to 3x3~mm$^{2}$. It consists of an array of p-n junction
pixels biased above the breakdown voltage. Each pixel is passively
quenched with an external resistor. Its response is a fixed amount of
charge for each impinging photon, hence not proportional to the photon energy.
The MPPC signal output is the sum of the charges of all pixels, which to first order is proportional to
the incident flux of photons.
The gain of the device ranges between $10^{5}$ and $10^{6}$.
The MPPC shows a high sensitivity in the 420 nm spectral region, with a photo-detection efficiency ranging between 25\% and 65\% depending on the pixel size. The typical low dark current ($<1$~$\mu$A), the low bias voltage ($\sim$70V) and the high gain largely simplify the readout electronics. 

In an earlier publication it has been shown that MPPC can be used to read out the blue light produced in an organic scintillator with a good light yield \cite{plastic}. The extension of that measurement to inorganic scintillators, with possible application to PET is the topic of this work. A first characterization of a basic detector unit is presented. A LSO crystal ($1\times 1\times 15$~mm$^{3}$ or $3\times 3\times 15$~mm$^{3}$) is read out by a MPPC with an active area of the same size active area. The energy resolution of one detector unit is measured at 511 keV. The time resolution of two detector units in coincidence is discussed. A first comparison between LSO (from Heilger Crystals) and LFS-7 (Lutetium Fine Silicate, developed by General Physics Institute, Moscow~\cite{LFS}) is presented. 


\section{The experimental setup}
This study is based on five samples of 1x1~mm$^{2}$ MPPCs (400 pixels) and five samples of 3x3~mm$^{2}$ MPPCs (3600 pixels).  The MPPCs are custom packaged and the active silicon is protected by a special plastic. The suggested operation voltage is 76~V and 69.9~V, respectively, with a spread of 0.1~V between the five pieces in each sample. The dark rate at 0.5 pixels is estimated to be 220-250~kHz and 3.2-3.3~MHz, respectively. The gain of the devices operated at nominal voltage is about $7\times 10^{5}$. \\
Two pairs of $1\times 1\times 15$~mm$^{3}$ and $3\times 3\times
15$~mm$^{3}$ LSO crystals and one pair of $3\times 3\times 15$~mm$^{3}$ LFS crystals are wrapped in a 2-mm thick Teflon layer. 
\begin{figure}[!t]
\centering
\includegraphics[width=0.7\textwidth]{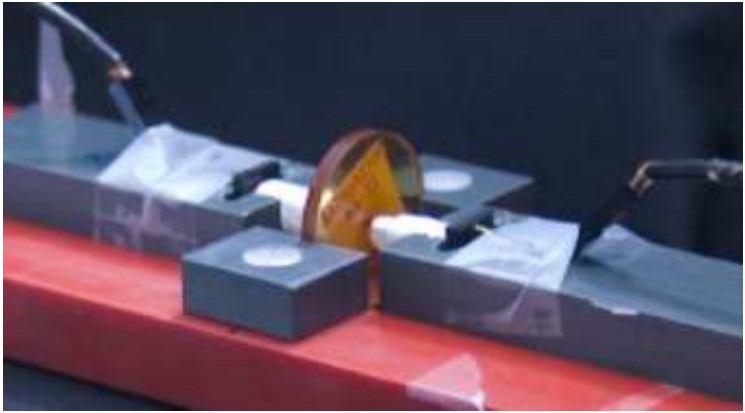}
\caption{\label{PT} Setup for the energy and time resolution measurements of LSO crystals. The scintillators (white) are directly read out by the MPPC and aligned with a $^{22}$Na source positioned in the middle. The setup allows a reproducibility of the measurements of 3$\%$ for the $3\times 3$~mm$^{2}$ detectors and $10\%$ for the $1\times 1$~mm$^{2}$ ones.}
\end{figure}
One end of the crystal is left free and coupled with optical grease to 
an MPPC of equal active area. 

The setup is shown in Fig.~\ref{PT}. Two special holders (gray) are machined to fix the photo-detector to the crystal front face. The precision in the relative alignment between the MPPC and the crystal is about $0.3$~mm. The holders are positioned face to face on a rail (red) on either side of the $^{22}$Na source (orange disk). The setup has one degree of freedom which allows to vary the distance between the two detectors as well as the distance between the detectors and the $^{22}Na$ source.\\
The signals from the MPPCs are digitized without any
amplification. For the energy resolution measurement the integration is performed using a VME QDC Lecroy 1182, gated by a coincidence of the two MPPC signals generated using standard NIM logic. 
For the time resolution measurement the signals are digitized using a
4-GHz True-Analog Bandwidth oscilloscope (TDS7404B by Tektronix) and
stored for offline analysis. In this case the trigger coincidence is
generated internally in the oscilloscope. \\
In acquisition mode the oscilloscope provides a sampling rate of 20 GS/s. The time resolution of each signal is 100 ps as two channels are used at the same time. The acquisition rate is quite poor ($\sim$1 Hz), however storing the waveforms allows large flexibility in the subsequent studies.

\section{Energy resolution} 
\begin{figure}[!t]
  \centering
      {\includegraphics[width=0.48\textwidth]{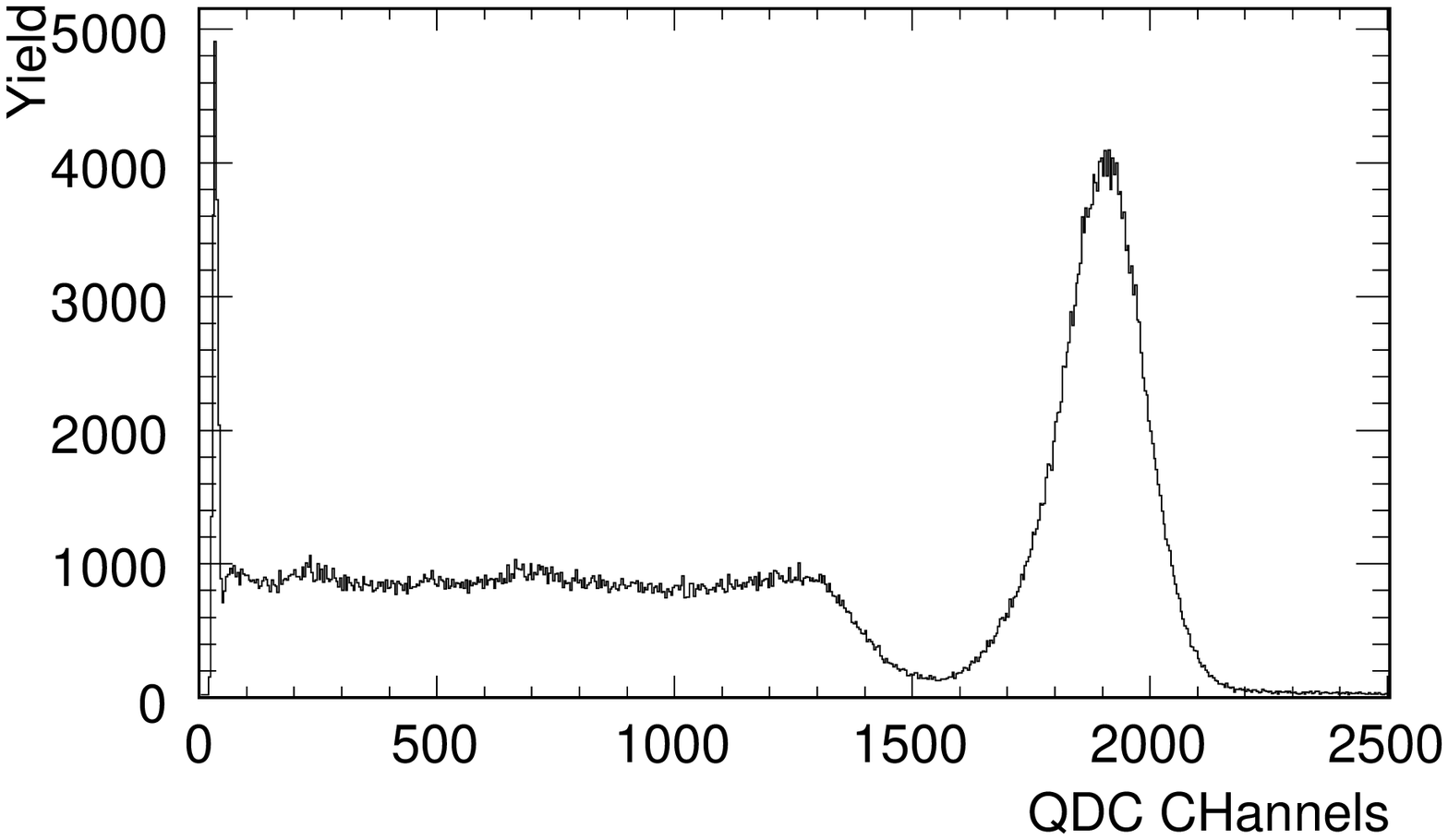}}
      {\includegraphics[width=0.48\textwidth]{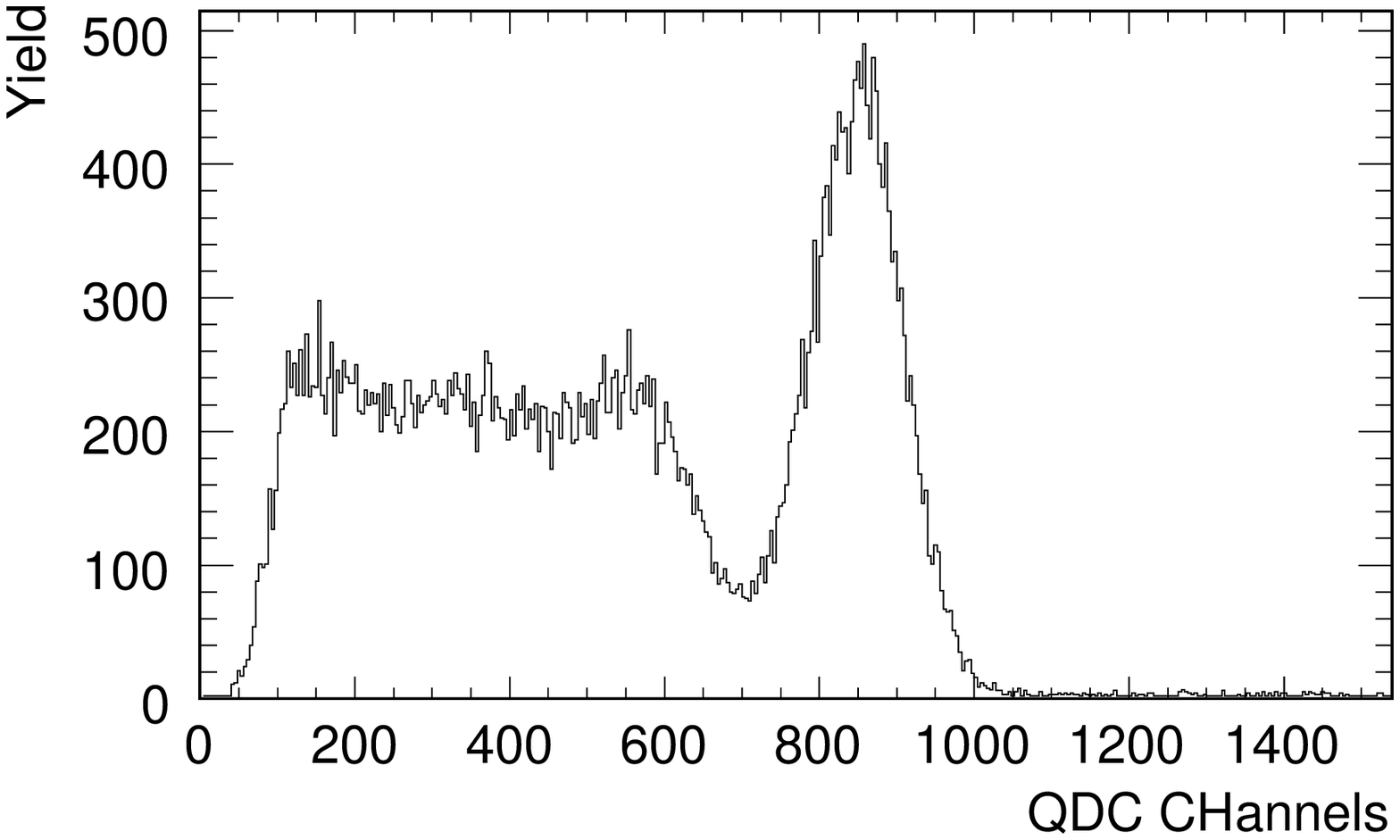}}
      \caption{\label{energy} Energy response to a $^{22}$Na source 
        (left) of $3\times 3\times 15$~mm$^{3}$ LSO crystal coupled with a
        $3\times 3$~mm$^{2}$ MPPC (3600 pixels), and (right) of $1\times 1\times 15$~mm$^{3}$ LSO crystal coupled with a $1\times 1$~mm$^{2}$ MPPC (400 pixels).}
\end{figure}

The energy spectrum of 511 keV photons measured with one detector is presented in Fig. \ref{energy}. The photo-electron peak is clearly separated from the energy continuum of Compton-scattered photons. The energy resolution of the detector is extracted using a Gaussian fit to the photo-electron peak. The source of systematic error are the alignment of the crystals with the source, the coupling with the photo-detectors and the wrapping. The systematic error is extimated repeating the measurement many times.  The FWHM/mean of the fit is quoted. 
An energy resolution  of $10.0 \% \pm 0.3\%(stat)\pm 1\% (sys)$ is
obtained for the $3\times 3\times 15$~mm$^{3}$ system
(Fig. \ref{energy}.a), while $14 \% \pm 0.4\%(stat) \pm 2\%(sys)$ is
measured with the $1\times 1\times 15$~mm$^{3}$  system
(Fig. \ref{energy}.b). The lower statistics of Fig. \ref{energy}.b
with respect to Fig. \ref{energy}.a is due to the reduced acceptance
of the $1\times 1\times 15$~mm$^{3}$  system. 
The rather large systematic uncertainties of the $1\times 1\times 15$~mm$^{3}$ system measurements are due to a still imperfect setup of the test system; improvements are possible, especially concerning the technical reproducibility and the crystal-MPPC coupling.

The measured energy resolution allows an efficient separation between the photo-electric peak and the Compton scattered events. It has been shown in  similar experiments \cite{dg3,Otte}  that the traditional SiPM (from CPTA and MEPHI) coupled to  a $3\times 3\times 15$ mm$^3$ crystal provides a resolution of $\sim$25-35\%, due to the poor photo-detection efficiency in the blue spectral region.
Furthermore, LSO crystals show $\sim$10\% energy resolution at 511~keV when read out by a traditional photomultiplier tube~\cite{Bill1} (LSO intrinsic energy resolution is $\sim$9\%~\cite{LSO1}). The results obtained indicate that the MPPC provides a energy resolution for PET application which is competitive with that of PMT with the advantage of an easy direct coupling to a small crystal.\\
The finite number of pixels of the MPPC causes its response to be non-linear at high photon fluxes\footnote{The photon flux is defined as the number of photons per mm$^{2}$ per ns. A Silicon PhotoMultiplier is in the linear region for fluxes of the order of $\frac{1}{\epsilon A_{p} \tau_{r}}\sim 400-500$~ns$^{-1}$~mm$^{-2}$ - $A_{p}$ being the area of a single pixel, $\tau_{r}$ its recovery time and $\epsilon$ the detection efficiency. }.  
The effect of the non-linearity of MPPC on the energy scale is investigated measuring the response of this system to $^{137}$Cs (611 keV), $^{122}$Ba (80 keV, 320 keV) as well as $^{22}$Na(511 keV).
\begin{figure}[th]
  \centering
      {\includegraphics[width=0.6\textwidth]{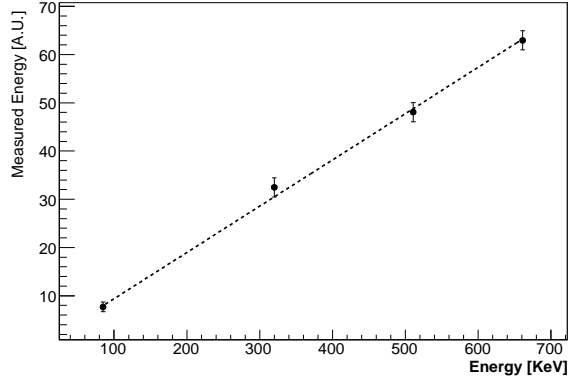}}
    \caption{\label{Linearity} Energy response to various gamma ray sources of a $3\times 3\times 15$~mm$^{3}$ LSO crystal coupled with a $3\times 3$~mm$^{2}$ MPPC (3600 pixels). The system shows a linear behavior.}
\end{figure}
Fig.~\ref{Linearity} shows a linear response in the region of interest, up to 611 keV. In fact the photon flux from the crystal is low, although their integral number is large. The photons are emitted in a wide time window, of the order of the decay time of the scintillator (40-50 ns), while the typical recovery time of the MPPC is only about 4~ns. The consequence is that the pixel recovers faster compared to the light emission rate and the saturation mechanism is strongly suppressed. 
\begin{figure}[!b]
  \centering
      {\includegraphics[width=0.6\textwidth]{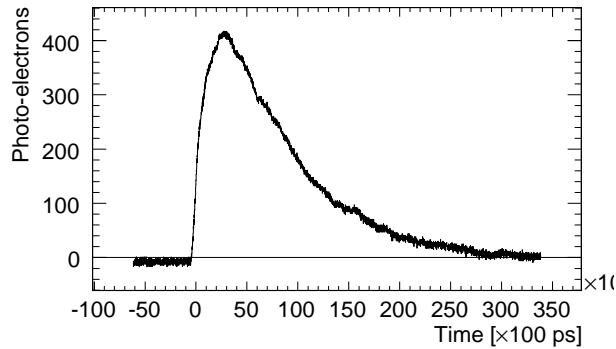}}
    \caption{\label{signal} Signal of a 511 keV photon detected (photoelectric effect) by a $3\times 3\times 15$~mm$^{3}$ LSO crystal coupled with a $3\times 3$~mm$^{2}$ MPPC (3600 pixels). The amplitude is shown in unit of a single photo-electron signal.}
\end{figure} 
The signal corresponding to the photo-electric interaction of a 511 keV photon in LSO is shown  in Fig.~\ref{signal}. 
The amplitude, rescaled to the single photo-electron size, gives an indicative order of magnitude of the time distribution of the detected photons\footnote{The integral is directly proportional to the total number of photo-electrons.}. The instantaneous amplitude never exceeds 500 photo-electrons. The probability that two or more photons are detected in the same pixel is hence minimal in this setup, as a maximum flux of 500 photo-electrons is distributed on 3600 pixels and on a total active area of $3\times 3$~mm$^{2}$.

\begin{figure}[!t]
  \centering
      {\includegraphics[width=0.6\textwidth]{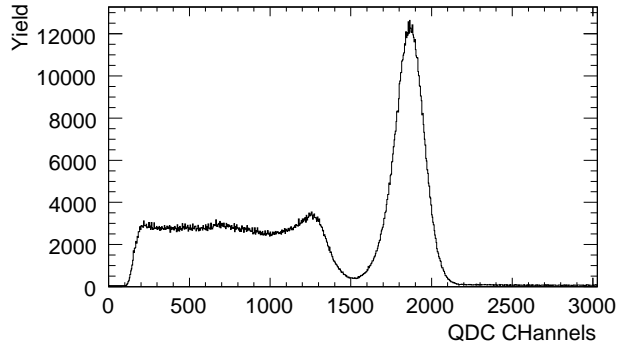}}
    \caption{\label{LFS} Energy response to a $^{22}$Na source of a $3\times 3\times 15$~mm$^{3}$ LFS crystal coupled with a $3\times 3$~mm$^{2}$ MPPC (3600 pixels). The obtained resolution is $11\%$, comparable within systematic uncertainties to the reaults obtained for the same-size LSO crystal.}
\end{figure}

The measurements were repeated for the $3\times 3\times 15$~mm$^{3}$ LFS crystal. It shows an energy resolution of $11\%$ (Fig. \ref{LFS}), comparable, within the systematic uncertainty, to the same-sized LSO crystal. 


\section{Time resolution}
The time resolution of the system is determined measuring the time difference between the two signals of two back to back scattered photons.
As an estimate of the time resolution the FWHM of the time difference distribution is taken. For the measurement of the signal timing a fixed-amplitude threshold is used in this study, instead of the constant fraction discriminator method, traditionally used in TOF-PET system. This is justified due to the fast response of the LSO crystals together with the fast rise time of the large photo-electron signal of the MPPCs, and significantly simplifies the readout electronics.  It requires the calibration of each detector cell to the same light yield which is easily achieved tuning the bias voltage of the MPPCs.
   
The two signals from the detector elements are directly sent to the inputs of the oscilloscope, where they are discriminated if above a tunable threshold. This threshold is kept at $\sim$4~mV (or $\sim$13-15 pixels). The minimum allowed threshold is constrained by the electronic noise level ($2.0\pm0.5$~mV corresponding to $10\pm1$~photo-electrons).
A coincidence is formed after the discrimination and used as trigger
to store the full signal waveform, starting considerably before the trigger time. The offline analysis is, hence, independent on the coincidence threshold.  

\begin{figure}[!th]
  \centering
      {\includegraphics[width=0.6\textwidth]{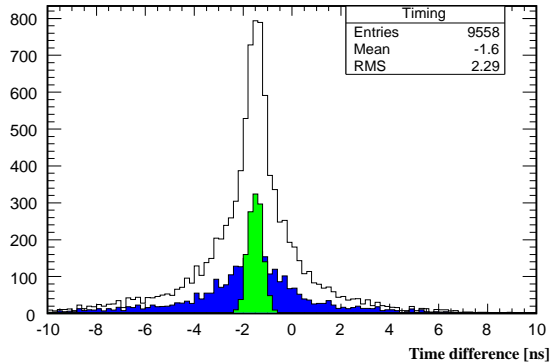}}
    \caption{\label{background}Time resolution of a system of two $3\times 3\times 15$~mm$^{3}$ LSO crystal read out directly by the MPPCs of the same size. The total sample, without cuts, is shown as a black line. The sharp signal peak (green) corresponds to the events whith energies near the 511 KeV photo-electric peak in both crystals. The blue background corresponds to the events in which one of the two photons looses energy by the Compton effect.}
\end{figure}

\begin{figure}[!b]
  \centering
      {\includegraphics[width=0.6\textwidth]{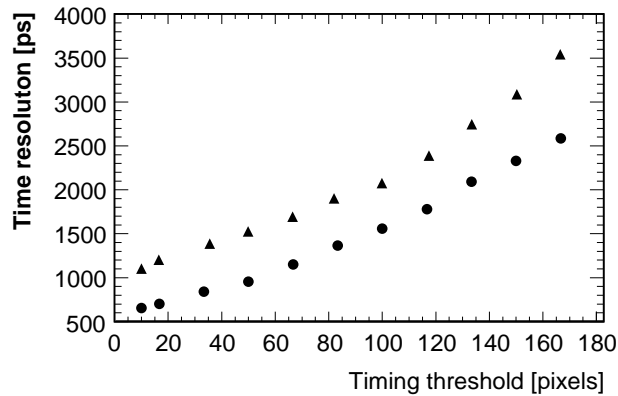}}
    \caption{\label{scanV}Time resolution as a function of the chosen fixed-amplitude threshold for a system of two $3\times 3\times 15$~mm$^{3}$ LSO crystals coupled to $3\times 3$~mm$^{2}$ MPPC 3600 pixels using fix-amplitude threshold. Results obtained with (points) and without (triangles) energy cut are shown.}
\end{figure}

The timing measurement is mainly influenced by the selection of the signals and the timing threshold, as it was previously shown in \cite{Kobe}.
Figure~\ref{background} illustrates the improvement of time resolution obtained when applying an energy cut of $\pm$1$\sigma$ around the photo-electric peak value.
When selecting only events with energies near the photo-electric peak value a sharp time difference distribution is observed (green). Its FWHM is measured to be $647\pm3 $~ps, estimated with a gaussian fit in the interval $\pm2 \sigma$ around the mean value. The main background of the measurement are events in which one or both photons undergo Compton scattering. The timing spread of these events is widely distributed and ruins the time resolution of the system (blue).  
This effect can be directly extrapolated from the  signal shapes observed in the oscilloscope. Photons which undergo Compton scattering are observed as signals of smaller amplitude and slope as compared to signals from photons depositing their full energy inside the crystals.
The influence of the chosen threshold on the time resolution is shown Fig.~\ref{scanV}.     
The time resolution degrades fast with increasing coincidence threshold, as the measurement becomes more sensitive to the variation of the rise time of the signals. The improvement in the observed time resolution when selecting events from the photo-electric peak is almost a factor of 2.\\
In a similar study of the timing properties of LSO crystals readout by photo-multipliers \cite{Bill2} a time resolution of 475~ps is quoted. The result obtained here using MPPCs is somewhat worse but of the same order of magnitude. However, the coincidence threshold would have to be lowered to below the amplitude of a signal corresponding to a single photo-electron, in order to fully benefit from the fast intrinsic time resolution of the MPPC. This region is outside the dynamical range of the current instrumentation, but will be analysed after the improvements of the present setup.
 
\section {Conclusion}
This study demonstrates that the technological achievements in the photo-detector development open the possibility of the design of new generation PET machines, with improved space resolution and sensitivity.    The energy resolution of a $3\times 3 \times 15$~mm$^3$ LSO crystal, directly read out by a MPPC with an active area of the same size, reaches $10\%$ FWHM and the timing resolution $650$~ps. Sligthly worse results - $\sim 14\%$ energy resolution - are obtained with the $1\times 1 $~mm$^2$ crystals and photo-detectors, mainly due to systematic effects in the alignment of the setup. The study thus shows that  MPPC are competitive with traditional photo-detectors currently used in Positron Emission Tomography. In addition, they would allow a significative simplification of the technological design and of the readout electronics. 

More systematic studies of the $1\times 1 $~mm$^2$ MPPC as well as of the energy and timing behaviour of the LFS crystals are needed. 
The next step is
the construction of a small size prototype consisting of two matrices
of 6x6 crystals each, for a total of 72 channels with individual
readout. The purposes of such prototypes are manifold. First the
channel-to-channel  homogeneity and reproducibility of the concept has
to be tested. A solution for the necessary multi-channel readout has
to be found, in order to make thi PET concept scalable to a larger prototype for commercial
use. Furthermore, the calibration and monitoring requirements of a multi-channel
detector need to be addressed as well as the stability of
operation. Last not least, a small prototype will give the
opportunity to test the improvement of time resolution in the 2D and
maybe 3D spatial reconstruction of a non-point-like radioactive
source.\\

\section*{Acknowledgment}
This work is supported by the Helmholtz-Nachwuchsgruppen grant VH-NG-206 and the BMBF, grant no. 05HS6VH1.
We thank Hamamatsu, which kindly provided us with the tested samples of MPPC. We also thank V. Koslov and A. Terkulov from LPI for providing the LFS crystals and for their help and useful suggestions during tests. Finally we thank  P. Smirnov for his support in establishing the laboratory setup.


\begin{thebibliography}{10}

\bibitem{PET1} P. Valk, D. Bailey, D. Townsend, M. Maisey, \textit{Positron Emission Tomography},Springer (2003)

\bibitem{PET2} J.L. Humm, A. Roszenfeld, A. Del Guerra, \textit{From PET detectors to PET Scanners}, European Journal of Nuclear Medicine, num. 11, vol. 30, 1574 (2003)

   \bibitem{Bill1}  W.W. Moses, \textit{Recent advances and future advances in time-of-flight PET},
 to be published in Nucl.Instr.Methods (2007).
 
   \bibitem{Bill2}  W.W.Moses, \textit{Prospects for Time-of-Flight PET using LSO Scintillator},
 IEEE Transactions on Nuclear Science \textbf{NS-46}, 474 (1999). 

   \bibitem{Lecomte}  R.Lecomte, \textit{Technology challenges in small animal PET imaging},
 Nucl.Instr.Methods  \textbf{A527}, 157 (2004). 
 
\bibitem{Hama} Hamamatsu, \textit{Multi-Pixel Photon Counter, Datasheet}.

\bibitem{SIPM} P.Buzhan et al. \textit{Silicon Photomultiplier and its possible applications}, Nucl.Instr.Methods \textbf{
A504}, 48 (2003).

\bibitem{plastic} N.D'Ascenzo et al., \textit{Study of Micro Pixel Photon Counters for a high granularity scintillator based hadron calorimeter}, arxiv:0711.1287v1 (2007).

\bibitem{LFS} A.I.Zagumennyi,Yu.D.Zavartsev,S.A.Kutovoi.Patent US 7,132,060 PCT Filed:Mar.12, 2004.
 \bibitem{LSO1} M. Balcerzyk et al.,\textit{YSO, LSO, GSO and LGSO. A Study of Energy Resolution and Nonproportionality},IEEETrans.Nucl.Science \textbf{47}, 1319 (2000).
.  
\bibitem{dg3} D.J. Herbert et al. \textit{Study of SiPM as a potential photo detector for scintillator readout},
 Nucl.Instr.Methods \textbf{A567}, 356 (2006).

\bibitem{Otte} A.N.Otte et al. \textit{A test of silicon photomultipliers as readout for PET}, Nucl.Instr.Methods \textbf{A545}, 705 (2005).

\bibitem{Kobe} N.D'Ascenzo \textit{Application of MPPC to the Positron Emission Tomography},
 PoS(PD07)006 (2007).

\end{thebibliography}
\end{document}